\documentstyle[12pt,epsf]{article}
\newcommand{\la}[1]{\label{#1}}
\newcommand{\be}{\begin{equation}}
\newcommand{\ee}{\end{equation}}
\newcommand{\ba}{\begin{eqnarray}}
\newcommand{\ea}{\end{eqnarray}}
\newcommand{\bi}{\begin{itemize}}
\newcommand{\ei}{\end{itemize}}
\newcommand{\rmi}[1]{{\mbox{\scriptsize #1}}}

\newcommand{\nr}[1]{(\ref{#1})}

\newcommand{\nn}{\nonumber \\}
\newcommand{\fr}[2]{{\frac{#1}{#2}}}
\newcommand{\msbar}{\overline{\mbox{\rm MS}}}

\newcommand{\dig}[1]{$\times10^{- #1}$}
\newcommand{\bmu}{\bar\mu}
\newcommand{\eq}{eq.\,}
\newcommand{\eqs}{eqs.\,}

\def\lsi{\raise0.3ex\hbox{$<$\kern-0.75em\raise-1.1ex\hbox{$\sim$}}}
\def\gsi{\raise0.3ex\hbox{$>$\kern-0.75em\raise-1.1ex\hbox{$\sim$}}}

\begin{document}

\begin{titlepage}
\begin{flushright}
CERN-TH/98-167\\
hep-lat/9805024\\
\end{flushright}
\begin{centering}
\vfill

{\bf INVERSE SYMMETRY BREAKING \\
WITH 4D LATTICE SIMULATIONS}
\vspace{0.8cm}

K. Jansen$^{\rm a}$\footnote{karl.jansen@cern.ch} and
M. Laine$^{\rm a,b}$\footnote{mikko.laine@cern.ch}  \\

\vspace{0.3cm}
{\em $^{\rm a}$Theory Division, CERN, CH-1211 Geneva 23,
Switzerland\\}
\vspace{0.3cm}
{\em $^{\rm b}$Department of Physics,
P.O. Box 9, 00014 University of Helsinki, Finland\\}
\vspace{0.3cm}

\vspace{0.7cm}
{\bf Abstract}

\end{centering}

\vspace{0.3cm}\noindent
According to resummed perturbation theory, certain scalar theories 
have a global symmetry, which is restored in the vacuum but is broken 
at high temperatures. Recently, this phenomenon has been studied with
4d finite temperature lattice simulations, and it has been suggested that 
the non-perturbative dynamics thus incorporated would hinder the transition. 
We have carried out another lattice study, for a theory with very small 
coupling constants. We find perfect compatibility with next-to-leading 
order resummed perturbation theory, and demonstrate that ``inverse'' 
symmetry breaking can indeed take place at high temperatures.
\vfill

\noindent
CERN-TH/98-167\\
May 1998\\

\vfill

\end{titlepage}

\section{Introduction}

Inverse symmetry breaking is a phenomenon where a (global) symmetry
is restored in the vacuum, but is broken at high temperatures. This 
is in contrast to the usual behaviour, where symmetries may be broken
at low temperatures but get restored at high ones. Originally,
inverse symmetry breaking was shown to be possible in O(N)$\times$O(M) 
scalar theories~\cite{w}, but something analogous might happen
in more realistic cases as well. Indeed, several potential
cosmological consequences have been discussed
(see, e.g., \cite{dvali} and references therein).

Recently, there has been some interest in studying inverse 
symmetry breaking with lattice simulations~\cite{b3d,b4d}.
This interest is based on controversial statements
existing in the literature,  
concerning the possibility of inverse symmetry breaking when
non-perturbative effects are taken into account (see, e.g., 
\cite{blo}--\cite{others} and references therein). The result of
the simulations, both in three dimensions (3d)~\cite{b3d}
and four dimensions (4d)~\cite{b4d}, was that no sign of
inverse symmetry breaking was seen, even though perturbatively 
it was supposed to take place.

The purpose of this paper is to study inverse symmetry
breaking with 4d finite temperature lattice simulations, 
in a theory that is coupled weakly enough for   
perturbation theory to work. We do observe clear signs
of inverse symmetry breaking. We also discuss the general
way in which non-perturbative effects may manifest 
themselves in scalar theories of the studied type, by constructing 
the relevant dimensionally reduced 3d effective theory.

To be more specific, consider the Euclidian scalar Lagrangian
\ba
{\cal L_\rmi{cont}} & = & 
\fr12 (\partial_\mu \phi_1)^T (\partial_\mu \phi_1)+
\fr12 (\partial_\mu \phi_2)^2+
\fr12 m_1^2\phi_1^T\phi_1 + \fr12 m_2^2 \phi_2^2 \nn
& + & \frac{1}{24} \lambda_1(\phi_1^T\phi_1)^2 +
\frac{1}{24} \lambda_2 \phi_2^4 +
\fr14\gamma \phi_1^T\phi_1 \phi_2^2, \la{contaction}
\ea
where $\phi_1$ is an O(N) symmetric real vector (we take N$=4$) and 
$\phi_2$ is a $Z_2$ symmetric real scalar field. The symmetry groups
appearing here are not essential for the existence of inverse
symmetry breaking. In~\cite{b3d,b4d}, a case
where both fields are real scalars and 
the symmetry is $Z_2\times Z_2$ was considered,  
but we have chosen to take an O(4) field to undergo inverse symmetry
breaking, since the finite volume effects in the simulations are 
then somewhat easier to control (a $Z_2$-field in the broken phase
has problematic ``tunnelling correlations'', 
because of a barrier separating the two degenerate minima).

Consider now the allowed values of the coupling constant $\gamma$ 
in \eq\nr{contaction}. In principle, $\gamma$ can be negative, even 
though not by an arbitrarily large amount. This is because the stability
of the potential at zero temperature requires that $\lambda_1\gg 0,
\lambda_2\gg 0$,
$\lambda_1\lambda_2 \gg 9 \gamma^2$. However, within 
these requirements, it may happen that
$(N+2) \lambda_1+3 \gamma \ll 0$ or $\lambda_2+N \gamma \ll 0$.
These combinations of the coupling constants turn out to 
multiply the temperature squared in the effective 
finite temperature mass parameters of $\phi_1$ and $\phi_2$, 
respectively. Thus, at high temperatures, one of 
the symmetries (but not both of them) may get broken, 
even if both symmetries are restored at zero temperature.
This is called inverse symmetry breaking. If a symmetry is 
broken already at $T=0$, one talks of ``symmetry non-restoration''.

The obvious way to establish the phenomenon of inverse symmetry
breaking would be to fix $m_1^2>0$ so that $\phi_1$ is in the symmetric 
phase at $T=0$, and then to increase the temperature. 
Equivalently, one can {\em fix} $T$
{\em and tune} $m_1^2$: it is then sufficient to show that there is a 
phase transition at some critical value $m_{1,c}^2>0$. 
Although we will use 4d finite temperature simulations to
study inverse symmetry breaking, we will use 
a 3d effective field theory to predict the values
of the parameters where this phenomenon should occur. 

In the next section, we derive the perturbative estimates
for inverse symmetry breaking in 
the scalar theory of \eq\nr{contaction} in some more detail, using
the method of dimensional reduction and 3d effective
field theories~\cite{dr}. This allows us to implement the
resummations needed at finite temperature in a transparent
way and, furthermore, to see what kind of non-perturbative effects 
there can be. In sect.~3, we study the theory in \eq\nr{contaction} 
with 4d finite temperature lattice simulations. The results of the 
simulations, a comparison with perturbation theory, 
and our conclusions are in sect.~4.

\section{The 3d effective theory: \\ perturbative 
and non-perturbative results} \la{pert}

It is well known that perturbation theory at finite temperature
requires resummations. A convenient way to implement
the resummations is to construct an effective 3d field theory
for the zero Matsubara modes. The construction of the 3d theory
is purely perturbative, as only massive degrees of freedom are
integrated out. The non-perturbative dynamics of the system
is contained in the final effective 3d theory; it can thus
be studied in an economic way with this approach.

In more concrete terms, the purpose of this section is to derive
the coefficients $c_1,c_2$ in the expression 
\be
\frac{m_1^2}{T_c^2} = c_1 |\gamma| +c_2 |\gamma|^{3/2} 
+ c_3 |\gamma|^2 + \ldots,
\la{expansion}
\ee
where $T_c$ is the temperature where inverse symmetry breaking takes
place, and we assumed that parametrically $\lambda_1,\lambda_2\sim |\gamma|$,
and $m_2^2\sim 0$. 
Moreover, we review why,  in this expression, 
non-perturbative effects only affect
the coefficient $c_3$ and the higher-order terms.
In general, non-perturbative effects also affect 
the order of the phase transition (the transition is of second order, 
as we will see), but the existence of the transition itself 
(i.e. the coefficients $c_1,c_2$) can be deduced purely perturbatively.

The general method of dimensional reduction, its accuracy, and 
generic rules for its application~\cite{generic}, have been discussed 
extensively in the literature~(see also \cite{jp,bn}, and~\cite{ms} 
for a review). Hence we discuss only the results here. 

In order to
derive the coefficients $c_1,c_2$, it is sufficient to perform 
dimensional reduction at the tree-level for the coupling constants and
at the 1-loop level for the mass parameters. In the body of the text
we work at this order, but in the appendix we give also the 
next corrections (of the relative order ${\cal O}(|\gamma|)$
with respect to the leading terms). 
This serves to show how the $\msbar$ scheme 
scale parameter appearing in the 
zero temperature parameters gets fixed. Moreover, 
the 2-loop corrections in the mass parameters are needed if 
the non-perturbative constant
$c_3$ were to be determined with lattice simulations in 
the 3d effective theory.

The first step of dimensional reduction is the integration out
of non-zero Matsubara modes. After rescaling the fields squared 
and coupling constants by $T$, the final effective 3d theory is
\ba
S_\rmi{eff} & = & \int d^3x \biggl\{  
\fr12 (\partial_i \phi_1)^T (\partial_i \phi_1)+
\fr12 (\partial_i \phi_2)^2+
\fr12 m_{1,3}^2\phi_1^T\phi_1 + \fr12 m_{2,3}^2 \phi_2^2 \nn
& + & \frac{1}{24} \lambda_{1,3}(\phi_1^T\phi_1)^2 +
\frac{1}{24} \lambda_{2,3} \phi_2^4 +
\fr14\gamma_3 \phi_1^T\phi_1 \phi_2^2
\biggr\}. \la{draction}
\ea
To this order, the fields appearing are just the same as the 
zero Matsubara components of the 4d fields, apart from the trivial 
rescaling with $T$. To leading meaningful order, the expressions
for the parameters appearing in \eq\nr{draction} are
\ba
\lambda_{1,3} & = & T \lambda_1, \quad
\lambda_{2,3} = T \lambda_2, \quad
\gamma_3 = T \gamma, \\
m_{1,3}^2 & = & 
m_1^2 +  \frac{T^2}{24}
\biggl(\frac{N+2}{3} \lambda_{1}+\gamma \biggr), \quad
m_{2,3}^2 =  
m_2^2  + \frac{T^2}{24}
\biggl(\lambda_{2}+N \gamma \biggr), \la{m23} \la{drparams}
\ea
where the parameters on the RHS mean renormalized
$\msbar$ scheme parameters at some scale $\sim T$
(to be more precise, see the appendix).

Near the symmetry breaking phase transition, 
the theory in \eq\nr{draction} can be further simplified. 
Indeed, recall that we have chosen it to be the 
O(N) field $\phi_1$ that undergoes inverse symmetry breaking:
\be
(N+2)\lambda_1+3\gamma \ll 0. \la{isb}
\ee
Around the critical temperature, $m_{1,3}^2\sim 0$
(more precisely, as we will see, $m_{1,3}^2\sim |\gamma|^{3/2} T^2$).
On the other hand, it follows from \eq\nr{isb} and  
the vacuum stability requirement 
$\lambda_1\lambda_2 \gg 9 \gamma^2$ that
\be
\lambda_2 + N \gamma \gg 0.
\ee
Hence the effective mass parameter $m_{2,3}^2$ is ``large'', 
$m_{2,3}^2\sim |\gamma|T^2\gg m_{1,3}^2$, and the 
heavy excitations corresponding 
to the field $\phi_2$ can be integrated out.

The final effective 3d theory after integrating out $\phi_2$
is of the form
\ba
S_\rmi{eff} & = & \int d^3x \biggl\{  
\fr12 (\partial_i \phi_1)^T (\partial_i \phi_1)+
\fr12 \bar m_{1,3}^2\phi_1^T\phi_1 +
\frac{1}{24} \bar\lambda_{1,3}(\phi_1^T\phi_1)^2 \biggr\}. \la{3daction}
\ea
The parameters that appear are 
\ba
\bar \lambda_{1,3} & = & 
\lambda_{1,3}, \\
\bar m_{1,3}^2 & = & 
m_1^2 + \frac{T^2}{24}\biggl(\frac{N+2}{3}\lambda_{1}+\gamma \biggr)
-\frac{1}{8\pi} \gamma_3 m_{2,3}. \la{m13b}
\ea
The next corrections are given in the appendix.

Using these expressions, we can discuss the perturbative
(and non-perturbative)  
predictions for inverse symmetry breaking. To be
specific, let us fix the parameter values. To safely satisfy \eq\nr{isb}
and the vacuum stability requirement, we choose
\be
\gamma = -|\gamma|,\quad
\lambda_1 = \frac{3}{2(N+2)}|\gamma|,\quad
\lambda_2 = 10 (N+2) |\gamma|. \la{choice}
\ee
These refer to the $\msbar$ scheme parameters at some scale $\bmu_0$. 
Note that even though they are parametrically of the same order of magnitude, 
the coupling constants in \eq\nr{choice} are numerically widely different.
In particular, $m_{2,3}^2$ involving $\lambda_2$ 
is ``large'', and therefore the first
correction it induces in $\bar m_{1,3}^2$
(the last term in \eq\nr{m13b}) is also significant. 
However, further corrections are relatively much smaller, 
by ${\cal O}(\gamma_3/m_{2,3})$.  
To further simplify matters, let us take $m_2^2\approx 0$.

{\bf The phase structure and critical temperatures}. 
We have derived above the effective 3d theory in \eq\nr{3daction}, 
and argued that it describes the infrared
properties of the system non-perturbatively 
around the point $\bar m_{1,3}^2\sim 0$.
What does this imply? The non-perturbative properties of the 
theory in \eq\nr{3daction} are of course well known: the theory
has a second order phase transition into a broken phase when
$\bar m_{1,3}^2$ becomes negative. The exact value of the 
$\msbar$ scheme parameter $\bar m_{1,3}^2$ at the phase 
transition point could be determined with lattice simulations, 
but this is not important for us here. It is sufficient to 
know that, since the coupling $\bar \lambda_{1,3}$ is 
dimensionful, the result can only be of the form 
$\bar m_{1,3}^{2(\rmi{crit})} = (c/16\pi^2) \bar\lambda_{1,3}^2$, 
where $c\sim 1$ ($c$ could also be negative). 
The factor $1/(16\pi^2)$ arises from
typical 2-loop order (see the appendix) contributions to $c$, cancelling the 
scale dependence of $\bar m_{1,3}^{2}$.   

Using now the expression in \eq\nr{m13b}, the value 
of $\bar m_{1,3}^{2(\rmi{crit})}$ can be converted to the critical 
temperature. It is seen that the non-perturbative constant
$c$ only contributes to $c_3$ in \eq\nr{expansion}.

Thus, the constants $c_1,c_2$ in \eq\nr{expansion}
can be derived simply by equating \eq\nr{m13b} 
with zero. This corresponds to the next-to-leading
order accuracy for critical temperature discussed in
general in~\cite{arnold}, and in the present context in~\cite{blo}.
Using \eq\nr{choice}, $\bar m_{1,3}^2$ becomes
\be
\bar m_{1,3}^2 \approx m_1^2 +T^2 \biggl[- \frac{|\gamma|}{48}
+\frac{|\gamma|^{3/2}}{8\pi}\sqrt{\frac{20+9N}{24}}\biggr]. \la{1lm13}
\ee
To have symmetry breaking, the coefficient of $T^2$ must
be negative, so that we need, for $N=4$, 
\be
|\gamma|^{1/2} \ll \frac{\pi}{6}\sqrt{\frac{24}{20+9N}}=0.34.
\ee
To satisfy this requirement, we fix
in practice $|\gamma|^{1/2}=1/6$. Then the critical 
temperature for inverse symmetry breaking is determined by 
\be
\frac{m_1^2}{T_c^2} = \frac{|\gamma|}{48} - 
\frac{|\gamma|^{3/2}}{8\pi}\sqrt{\frac{20+9N}{24}} \approx 3\times 10^{-4}.
\la{tc}
\ee

{\bf The scalar field expectation value.}
Besides the critical temperature itself, we would also like to know 
the behaviour of $\langle\phi_1\rangle$ around $T_c$. It turns out
that a good enough estimate can already be obtained with the tree-level
result in the effective theory of \eq\nr{3daction}:
\be
\langle\phi_1\rangle = \sqrt{-\bar m_{1,3}^2/\bar \lambda_{1,3}}.
\la{1lphi}
\ee
We have also computed the 2-loop effective potential 
in the theory of \eq\nr{3daction}, as can be easily done
(then one also has to use the 2-loop parameters in the appendix), 
but numerically the effects found are very small for the small
coupling constants used, typically $\sim 1\%$.

\section{Finite temperature 4d lattice simulations} \la{latt}

We now turn to 4d finite temperature lattice simulations, 
and attempt to verify \linebreak 
\eqs\nr{tc}, \nr{1lphi}.
As a first point, let us discuss the renormalization of the 
theory. 

As mentioned above, the coupling constants considered
are very small. Thus perturbation
theory sensitive only to ultraviolet (UV) degrees of freedom is
well convergent. As a consequence, the theory can be renormalized
in perturbation theory. 

To be more specific about the renormalization, recall that
in sect.~\ref{pert} we have param\-etrized the theory by 
couplings defined in the $\msbar$ scheme
at an arbitrary scale $\bmu_0$.
In the simulations, in contrast, one uses the 
lattice regularization. To compute the relation between the schemes 
is a standard exercise (it is analogous to computing the relation
between $\Lambda_\rmi{QCD}$ in the continuum and on the lattice~\cite{lamqcd}, 
but much simpler as this is a scalar theory).  
One computes a set of 1-loop graphs and requires that the results
are the same in the $\msbar$ and lattice regularization schemes. 
At the 1-loop level, the result is as follows. Let
\be
L_a(\bmu) \equiv \frac{1}{16\pi^2}\left[
2 \ln(a\bmu ) + 16\pi^2 r_1+1 \right].
\ee
Then the bare parameters
appearing in the lattice action are
\ba
m_{1B}^2 & = & m_1^2(\bmu_0) -\frac{(N+2)\lambda_1+3\gamma}{6} \frac{r_0}{a^2}
-\frac{(N+2)\lambda_1 m_1^2+3\gamma m_2^2}{6}  L_a(\bmu_0), \nn
m_{2B}^2 & = & m_2^2(\bmu_0) -\frac{\lambda_2+N \gamma}{2} \frac{r_0}{a^2}
-\frac{\lambda_2 m_2^2+N \gamma m_1^2}{2} L_a(\bmu_0), \nn
\lambda_{1B} & = & \lambda_1(\bmu_0) - \fr32 
\left(\frac{N+8}{9}\lambda_1^2+\gamma^2\right)
L_a(\bmu_0), \nn
\lambda_{2B} & = & \lambda_2(\bmu_0) - \fr32
\left(\lambda_2^2+N \gamma^2\right) L_a(\bmu_0), \nn
\gamma_B & = & \gamma(\bmu_0)-\fr12
\left(4\gamma^2+\frac{N+2}{3}\gamma\lambda_1+\gamma\lambda_2\right)
L_a(\bmu_0). \la{lattparams}
\ea 
These bare parameters are of course independent of $\bmu_0$.
The constants that appear are~\cite{lw}
\be
r_0=0.154\,933\,390,\quad r_1=-0.030\,345\,755.
\ee
The lattice spacing appearing here is determined by
\be
a = \frac{1}{N_t T}, \la{eq:a}
\ee
where $N_t$ is the temporal extent of the lattice.

When scaled into a dimensionless form by 
$\phi_i\to (\sqrt{\kappa_i}/a) \phi_i$,
the lattice action becomes
\ba
S & = &  -\kappa_1 \sum_{x,\mu} \phi_1^T(x+\hat\mu)\phi_1(x) 
- \kappa_2  \sum_{x,\mu} \phi_2(x+\hat\mu) \phi_2(x) 
+ \sum_x \phi_1^T(x)\phi_1(x) + \sum_x \phi_2^2(x) \nn
& + & \beta_1\sum_x [\phi_1^T(x)\phi_1(x)-1]^2 
+ \beta_2\sum_x [\phi_2^2(x)-1]^2 
+ \alpha\sum_x \phi_1^T(x)\phi_1(x) \phi_2^2(x). \la{betas}
\ea
Here the parameters are related to those in \eq\nr{lattparams} by
\ba
\kappa_i & = & \frac{6}{\lambda_{iB}}
\biggl[-\Bigl(4+\fr12 a^2 m_{iB}^2\Bigr)+ 
\sqrt{\Bigl(4+\fr12 a^2 m_{iB}^2\Bigr)^2+\fr13 \lambda_{iB}}
\biggr],\nn
\beta_i & = & \frac{1}{24}\lambda_{iB} \kappa_i^2, \quad
\alpha = \fr14\gamma_B \kappa_1\kappa_2.
\ea

The general philosophy of the simulations is now as follows. 
It is seen from \eq\nr{1lm13} that, for a fixed $m_1^2$, the O(N) 
symmetry is expected to be broken when $T$ is increased 
above some critical value. Equivalently, as mentioned,
one can fix $T$ and tune $m_1^2$: when $m_1^2$ is smaller than some
critical value, but still positive, there should be 
symmetry breaking. The latter
viewpoint is easier to implement in simulations since, according
to \eq\nr{eq:a}, the lattice spacing is then kept fixed.  As there is
only one temperature, we can also choose, for convenience, 
the arbitrary fixed parameter $\bmu_0$ to coincide with $T$.
Then, we just vary $m_1^2(\bmu_0)/T^2$.

\begin{table}[t]
\centering
\begin{tabular}{lllllll}
\hline
$m_1^2/T^2$ & $\kappa_1/10^{-1}$ & $\kappa_2/10^{-1}$ & 
$\beta_1/10^{-5}$ & $\beta_2/10^{-3}$ & $\alpha/10^{-4}$ &
lattices \\ \hline
\multicolumn{7}{c}{$N_t=2$} \\ \hline
5.0\dig{5} & 1.2497845 & 1.2570608 & 1.8185293 & 
4.7496966 & -4.4763402 & $\{12^3...36^3\}$ \\ 
  &  &  &  &  &  & $12^4...20^4$ \\
1.5\dig{4} & 1.2497806 & 1.2570608 & 1.8185179 & 
4.7496967 & -4.4763263 & $\{20^3,24^3\}$\\
3.0\dig{4} & 1.2497747 & 1.2570608 & 1.8185009 & 
4.7496968 & -4.4763053 & $\{20^3,24^3\}$\\ 
6.0\dig{4} & 1.2497630 & 1.2570609 & 1.8184668 & 
4.7496969 & -4.4762634 & $\{20^3,24^3\}$\\ 
1.0\dig{3} & 1.2497474 & 1.2570609 & 1.8184213 & 
4.7496972 & -4.4762076 & $\{20^3...28^3\}$\\ 
2.0\dig{3} & 1.2497083 & 1.2570610 & 1.8183077 & 
4.7496977 & -4.4760680 & $\{20^3,24^3\}$\\ 
3.0\dig{3} & 1.2496693 & 1.2570610 & 1.8181941 & 
4.7496982 & -4.4759283 & $\{20^3...28^3\}$\\ 
4.0\dig{3} & 1.2496302 & 1.2570611 & 1.8180804 & 
4.7496988 & -4.4757887 & $\{20^3,24^3\}$ \\ 
6.0\dig{3} & 1.2495522 & 1.2570613 & 1.8178532 & 
4.7496999 & -4.4755096 & $\{12^3...36^3\}$ \\ \hline
\multicolumn{7}{c}{$N_t=4$} \\ \hline
5.0\dig{5} & 1.2497859 & 1.2568207 & 1.8213989 & 
4.8442880 & -4.5055481 & $\{24^3\}$\\ 
6.0\dig{3} & 1.2497278 & 1.2568209 & 1.8212295 & 
4.8442891 & -4.5053391 & $\{24^3\}$\\ \hline
\end{tabular}
\caption[a]{\protect
The coupling constants and the lattice sizes used.
The numbers in the curly brackets indicate the
spatial volume; the temporal extent is $N_t$.
The notation $L_\rmi{min}^n...L_\rmi{max}^n$ means
that the lattice extent $L$ is increased in steps of 4
from $L_\rmi{min}$ to $L_\rmi{max}$.}
\la{tab:statistics}
\end{table}

The parameter values following from \eqs\nr{lattparams},
with coupling constants as in \eq\nr{choice}
and $m_2^2(\bmu_0)=0$, are shown 
in Table~\ref{tab:statistics} together with the lattice sizes used.
The numerical simulations have been performed at small values of 
$\beta_1$ and $\beta_2$ in \eq\nr{betas} where the cluster algorithm is not
very effective. We have chosen instead a hybrid 
over-relaxation algorithm \cite{algo}. 
For each point we performed 50K sweeps for thermalization and 100K
sweeps for
measurements. The errors were always computed from a blocking analysis 
searching for a plateau behaviour. 

\section{Results and conclusions}

\begin{figure}[t]
 
\vspace*{-1cm}
 
\centerline{ 
\epsfxsize=11cm\epsfbox{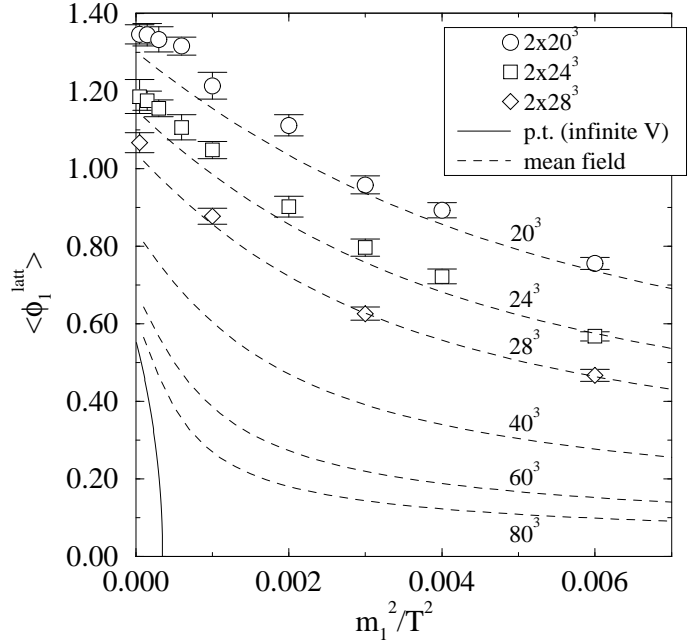}}
 
\vspace*{-5.5cm}
 
\caption[a]{\protect
A scan of lattice results, as a function of the continuum 
parameter $m_1^2/T^2$. The infinite volume perturbative continuum result, 
as well as finite volume mean field estimates, are also shown.
The lattice results are completely compatible with the perturbative ones.}
\la{fig:scan}
\end{figure}

\begin{figure}[t]

\vspace*{-1cm}
 
\centerline{ 
\epsfxsize=11cm\epsfbox{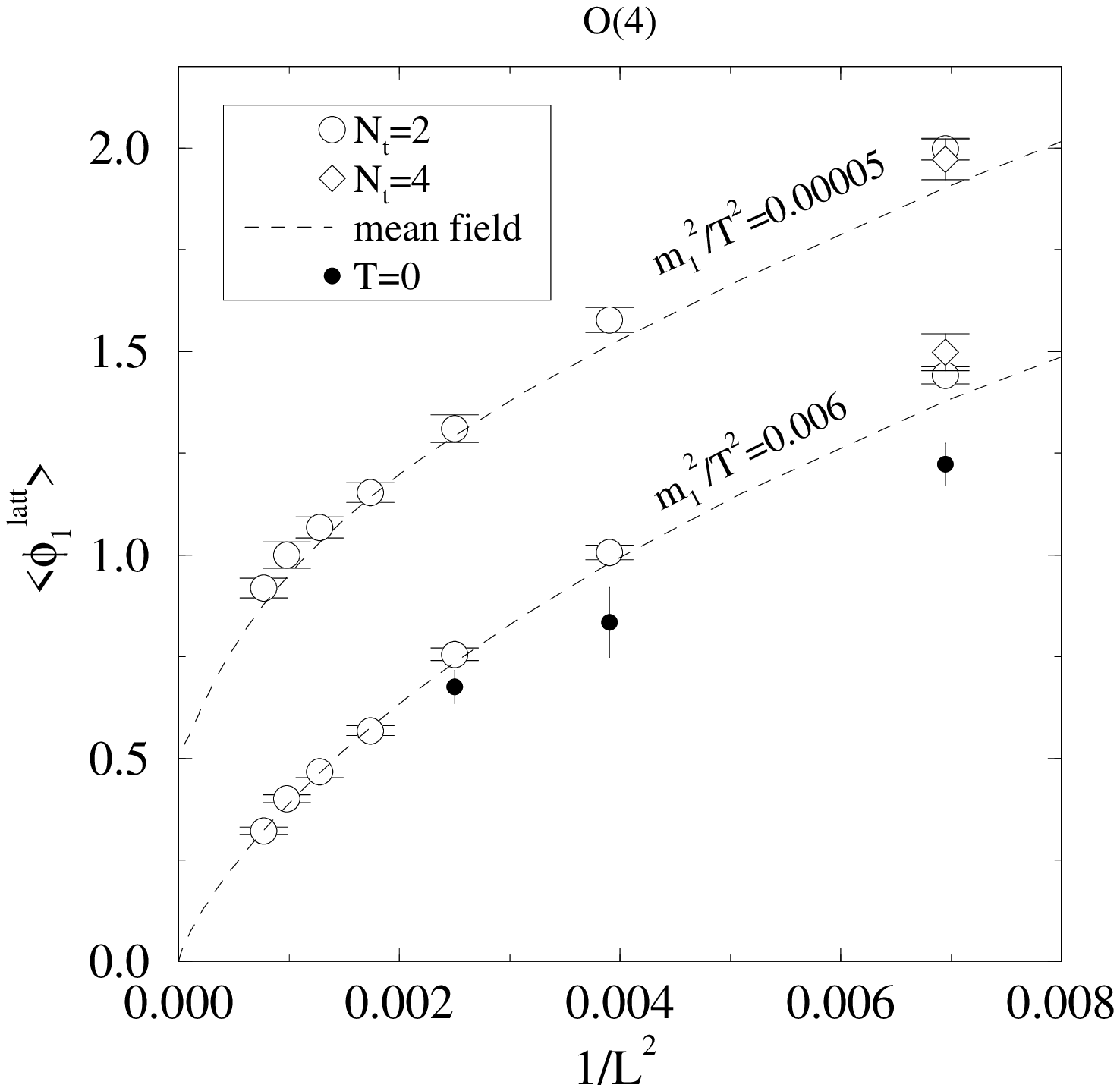}}
 
\vspace*{-5.5cm}

\caption[a]{\protect
The infinite volume limit for the O(4) field 
expectation value $\langle\phi_1\rangle$
at $m_1^2/T^2=0.00005,0.006$. Mean 
field estimates are also shown, as well as results from
a smaller lattice spacing.}
\la{fig:infvol}
\end{figure}

In Fig.~\ref{fig:scan}, we show a scan of continuum parameters, 
for different volumes. The field is shown in lattice units, 
$\phi_1^\rmi{latt}=(a/\sqrt{\kappa_1})\phi_1^\rmi{cont}$.
The infinite volume curve computed from the 2-loop 
effective potential in the theory of \eq\nr{3daction} is also shown
(however, as mentioned, the tree-level result could also be used, 
as the difference is very small, of order 1\%).

It can be seen that the lattice results differ from the 
continuum perturbative result by quite a significant amount 
for the volumes used. However, this can be easily understood.
Indeed, the finite volume mean field estimates obtained from
\be
\langle\phi_1\rangle \approx 
\frac{\int d\hat\phi_1 \hat\phi_1^4 
\exp[-\frac{1}{T} N^3 a^3 V_\rmi{eff}(\hat\phi_1)]}
{\int d\hat\phi_1 \hat\phi_1^3 
\exp[-\frac{1}{T} N^3 a^3 V_\rmi{eff}(\hat\phi_1)]},
\ee
where $V_\rmi{eff}(\hat\phi_1)$ is the continuum effective
potential, $\hat\phi_1$ denotes the radial component of~$\phi_1$, 
and the volume element of an O(4) vector is $d\phi_1\propto
\hat\phi_1^3d \hat\phi_1$, are also shown in the figure.
They are in perfect agreement with the lattice results. 

Next we take two points and look more precisely 
at the approach to the infinite volume limit. The 
results are shown in Fig.~\ref{fig:infvol}.
Again we find perfect compatibility with the
perturbative mean field estimates. For the parameter
value $m_1^2/T^2=0.00005$, there is inverse symmetry 
breaking, while for $m_1^2/T^2=0.006$, the symmetry
is restored in the infinite volume limit.

The results so far had been obtained with a single lattice
spacing, $N_t=2$. We have also made simulations with $N_t=4$.
The results are shown in Fig.~\ref{fig:infvol}, after
rescaling volumes and units so that they become
comparable with those at
$N_t=2$ (in other words, the volume $4\times 24^3$ is represented
at the same point as $2\times 12^3$, and the field has been 
normalized so that the data points correspond to the same 
continuum units). No lattice spacing dependence is seen
within the errorbars.

Finally, let us point out that, in principle, the existence of
inverse symmetry breaking could also be verified without any reference
to perturbative renormalization. In practice, though, this would 
require huge volumes for the weak couplings we have chosen. However, 
we have made a few zero temperature runs for the parameters 
corresponding to $m_1^2/T^2=0.00005$. The results are shown in 
Fig.~\ref{fig:infvol} with the filled circles. It is seen 
that the results for $\langle\phi_1\rangle$ are indeed smaller
than at finite temperature, and certainly consistent with an 
extrapolation to zero.

\smallskip

In conclusion, we have reiterated
what kind of non-perturbative effects there can be in finite 
temperature phase transitions in scalar theories. The existence
of the transition alone can be deduced with perturbation theory, 
while some of its properties (such as the order) are non-perturbative.
With explicit 4d finite temperature lattice simulations, we have verified
that the non-perturbative behaviour is indeed in perfect agreement 
with perturbative estimates, and inverse symmetry breaking takes place.

\section*{Acknowledgements}

Most of the simulations were carried out with a Cray C94 at the 
Center for Scientific Computing, Finland. We thank P. Hernandez,
K. Kajantie, C. Korthals Altes
and M. Shaposhnikov for discussions.

\section*{Appendix}

We give here the parameters appearing in the 
3d theories in \eqs\nr{draction} and \nr{3daction}  with more
precision than in the text.

Denoting $c= -0.348725$ and
\be
L_b(\bmu)=2\ln\frac{\bmu}{4\pi e^{-\gamma}T}\equiv 2\ln\frac{\bmu}{\bmu_T},
\ee
the $\msbar$ scheme renormalized
parameters appearing in \eq\nr{draction} are
\ba
\lambda_{1,3} & = & T \lambda_1(\bmu_T)
=T\biggl[\lambda_1(\bmu) - \frac{3}{32\pi^2}
\biggl(\frac{N+8}{9}\lambda_1^2+\gamma^2\biggr)L_b(\bmu)\biggr], \nn
\lambda_{2,3} & = & T \lambda_2(\bmu_T)
=T\biggl[\lambda_2(\bmu) - \frac{3}{32\pi^2}
\biggl(\lambda_2^2+N\gamma^2\biggr)L_b(\bmu)\biggr], \nn
\gamma_3 & = & T \gamma(\bmu_T)
=T\biggl[\gamma(\bmu)-\frac{1}{32\pi^2}
\biggl(4\gamma^2+\frac{N+2}{3}\gamma\lambda_1+\gamma\lambda_2\biggr)
L_b(\bmu)\biggr], \nn
m_{1,3}^2(\bmu) & = & 
m_1^2(\bmu) - \frac{1}{32\pi^2} \biggl(\frac{N+2}{3} \lambda_1 m_1^2+
\gamma m_2^2 \biggr) L_b(\bmu) + \frac{T}{24}
\biggl(\frac{N+2}{3} \lambda_{1,3}+\gamma_3 \biggr) \nn
& - & \frac{1}{32\pi^2} \biggl( 
\frac{N+2}{9}\lambda_{1,3}^2 + \gamma_3^2
\biggr)\biggl(\ln\frac{3T}{\bmu} + c \biggr), \nn
m_{2,3}^2(\bmu) & = & 
m_2^2(\bmu) - \frac{1}{32\pi^2} \biggl(\lambda_2 m_2^2+
N \gamma m_1^2 \biggr) L_b(\bmu) + \frac{T}{24}
\biggl(\lambda_{2,3}+N \gamma_3 \biggr) \nn
& - & \frac{1}{32\pi^2} \biggl( 
\fr13 \lambda_{2,3}^2 + N \gamma_3^2
\biggr)\biggl(\ln\frac{3T}{\bmu} + c \biggr). \la{2ldrparams}
\ea
The 3d coupling constants here are scale independent, whereas the 
mass parameters are not (due to the terms on the latter lines),
corresponding to the 2-loop UV-divergences remaining in the 
super-renormalizable 3d theory. When computing, for instance, the 2-loop
effective potential in the 3d theory, this scale dependence 
gets cancelled. Where not indicated explicitly, 
the parameters appearing in the expressions in \eqs\nr{2ldrparams}
are renormalized $\msbar$ scheme quantities.

The parameters appearing in the final effective theory in \eq\nr{3daction}
are 
\ba
\bar \lambda_{1,3} & = & 
\lambda_{1,3} -\frac{3}{16\pi}\frac{\gamma_3^2}{m_{2,3}}, \\
\bar m_{1,3}^2(\bmu) & = & 
m_1^2(\bmu_T) + \frac{T}{24}\biggl(\frac{N+2}{3}\lambda_{1,3}+\gamma_3\biggr)
-\frac{1}{8\pi} \gamma_3 m_{2,3} \nn
& + & \frac{1}{32\pi^2} \biggl[
\fr14 \lambda_{2,3}\gamma_3 -\gamma_3^2 
\biggl(\ln\frac{3T}{2 m_{2,3}} + c + \fr12 \biggr)
-\frac{N+2}{9} \bar \lambda_{1,3}^2 
\biggl(\ln\frac{3T}{\bmu}+c\biggr)\biggr]. \la{2l3dparams}
\ea
For $\bar m_{1,3}^2$, the first line is the 1-loop
expression and the second line the 2-loop contribution.
In these formulas, 
it is enough to use the 1-loop part of $m_{2,3}^2$,
given in \eq\nr{m23}.

\end{document}